\documentclass[adp,fleqn]{w-art}
\usepackage{times}
\usepackage{w-thm}
\theoremstyle{plain}

\theoremstyle{definition}

\usepackage[]{graphicx}
\chardef\bslash=`\\ 

\begin{document}
\DOIsuffix{theDOIsuffix}
\Volume{12}
\Issue{1}
\Copyrightissue{01}
\Month{01}
\Year{2007}
\pagespan{1}{}
\Receiveddate{8 May 2007}



\title[FCS and FT]{Full Counting Statistics and Field Theory}


\author[Yuli V. Nazarov ]{Yuli V. Nazarov 
     } 
\address{Kavli Institute of NanoScience, Delft University of
Technology, 2628 CJ Delft, The Netherlands}


\begin{abstract}
  We review the relations between the full counting statistics
  and the field theory of electric circuits. We demonstrate that
  for large conductances the counting statistics is determined
  by non-trivial saddle-point of the field. Coulomb 
 effects in this limit are presented as quantum corrections 
 that can stongly renormalize the action at low energies.

\end{abstract}
\maketitle
\section{Introduction}
The concept of Full Counting Statistics has been introduced in
very early days of quantum transport.
That time, nobody ever thought of such an abstract and complicated
problem as the evaluating and measuring the higher cumulants of
electronic noise. The research has been driven by pure curiosity
and has resulted in compact and deep Levitov formula.\cite{Lev93} 
The significance of this contribution has been underappreciated
for a number of years. The direct experimental verification
seemed to be out of question, while the theorists opted to address
more primitive and standard problems.

Nowadays Full Counting Statistics is a reasonably established 
field attracting attention of many, although the peak of interest 
is probably in the past.\cite{QuantumNoise}
There are beautiful experiments where the
higher cumulants have been measured \cite{Reulet} and even single electron transfers
have been actually counted.\cite{ElectronsCounted} 
Theoretically, the Full Counting Statistics
has been evaluated for virtually any important
electron transport system including even so generic as 
Anderson impurity model.\cite{AndersonFCS} 

Still it remains underappreciated that
the Full Counting Statistics is eventually not about 
the marginal deviations of electric currents.
Let as draw a parallel with general relativity. The general
relativity is not an art to calculate ridiculously small corrections
to Newton's law, although only those can be verified experimentally.
The general relativity brings us true knowledge about the Universe.
Similar to that, Full Counting Statistics lies near the heart
of quantum theory of electricity and is in fact an indispensible
element for this.

The present contribution aims to explicate the link between
Full Counting Statistics and quantum field theory of electric circuits.
The most established example of such theory is the quantum theory
of a superconducting Josephson junction in a dissipative 
electromagnetic environment \cite{AES} that is readily reduced
to a single-variable field theory for the superconducting
phase across the junction \cite{Grabert1}.
The action of the same type governs
Coulomb blockade phenomena in non-superconducting 
systems.\cite{Ave91,ASI,Grabert2}
The physics of one-dimensional interacting electrons
in the framework of Luttinger model is often reduced 
to similar schemes \cite{Kane92,Grabert3}, where the variable is a drop
of a phase over the barrier present in the one-dimensional
setup and can be associated with the voltage drop at the barrier.

The structure of the article is as follows. 
We start (Sections 2,3) by formulating a general quantum theory of a simplest
electric circuit and see the need and advantage of the FCS in this
respect. We show that the classical limit of the
field theory is not trivial as far as FCS is concerned and obtain 
the FCS at current bias (Section 4). Quantum effects at big 
conductances (as compared with the conductance quantum $G_Q \equiv e^2/2\pi \hbar$)
can be incorporated by the renormalization of the action 
parameters. This is frequently the case in field theories \cite{renormalization}
We perform the renormalization procedure explicitly in Section 5
for a quantum contact in series with an Ohmic one and
for a set of quantum contacts connected to a single node
(Section 6).

The concrete results revisited here were first published in
\cite{Beenakker},\cite{Kindermann}, and \cite{Bagrets}.

\section{Field theory}
Let us first start with elementary electric circuit theory
and reason the quantum extension of it.
A circuit is made of three archetypal elements: terminals,
connectors, and nodes. The voltage is fixed in terminals.
A connector is characterized by its $I-V$ characteristics:
\begin{equation}
I(t)\equiv I(t;\{V(t)\})
\end{equation}
where we have assumed general relation between the voltage
and current so that current at the time moment $t$ depends
on time-dependent voltage at all (previous) time moments, $\{V(t)\}$.
A simplest circuit contains two connectors ($A$ and $B$) 
in series, so that a single node and two terminals. (Fig. 1a)
Connecting elements in this way brings about an extra variable
:Voltage $V_1(t)$ in the node.
In the elementary circuit theory under consideration, 
this voltage  $V_1(t)$ is determined from the current conservation in the
node,
\begin{equation}
I_{A}(t;\{V_1(t)\})=I_{B}(t;\{V(t)-V_1(t)\}).
\end{equation}
assuming the terminal voltages are fixed to $0$ and $V(t)$.
Once the voltage is determined, one finds the $I-V$ characteristics
of the whole circuit. Thereby, the full description of the
system naturally emerges from the two descriptions of the separate connectors.

\begin{figure}
\centerline{\includegraphics[width=12cm]{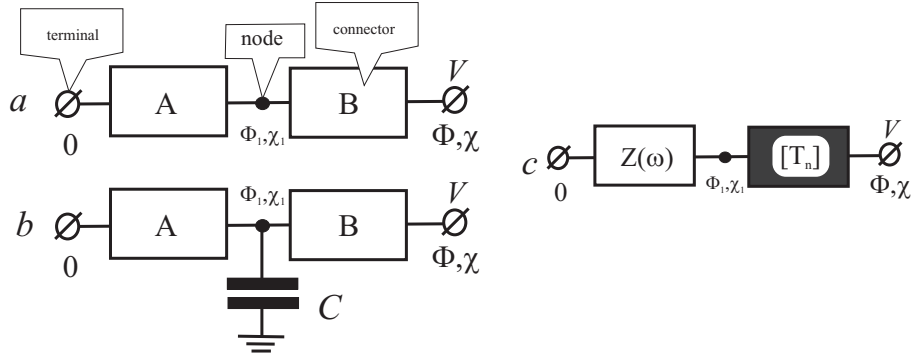}}
\caption{a. A simplest electric circuit consisting of
two terminals, two connectors {\it A,B} and one node,
gives rise to a quantum field theory for the field in the node.
b. Charging energy is represented as a capacitor
that provides high-frequency cut-off for the field. 
c. Setup studied in Sections 4,5: Quantum connector (characterized
by transmission eigenvalues $T_n$)
in series with Ohmic one (characterized by the frequency-dependent
impedance $Z(\omega)$). }
\label{circuit}
\end{figure}

Let us try to reason a quantum extension of this theory. First of
all, it is convenient to change from voltages to phases defined as
$\dot{\Phi}(t)=(e/\hbar)V(t)$. This allows to treat superconducting
and non-superconducting systems on equal footing. The phases(voltages)
of the terminals can be regarded as time-dependent external parameters
while the phase of the node becomes a real quantum variable, that is,
an operator $\hat{\Phi}_1(t)$. However, it is hardly an option 
to formulate the theory in operator formalism since finding
the classical correspondence becomes a formidable task. Rather,
we shall opt for Feynman-Vermon or Keldysh-action description 
of the system where all observables can be presented as path integrals 
over the time-dependent non-operator variable $\Phi(t)$. The price
to pay for this convenience is quite known: the variable "doubles".
The point is that the path integral should performed over two
parts of the Keldysh contour that correspond to coherent evolution 
of "kets" and "bra"'s. The variable in principle takes
different values $\Phi_1^+(t),\Phi_1^-(t)$ at these two contours.

We still want the description of the compound system under consideration to
emerge from the two descriptions of the separate connectors $A$ and $B$. 
What descriptions? Since the variable has "doubled", the
simple $I-V$ characteristics does not suffice. A connector
has to be described by something that depends on both
variables $\Phi_1^+(t),\Phi_1^-(t)$. It is also clear
that from all possible quantum variables of the connector
such description should involve only one: the operator
of electric current through the connector, $\hat{I}(t)$,
and present the reaction of the current on the variables 
$\Phi_1^+(t),\Phi_1^-(t)$. The proper description of
the connector is thus provided by the Feynman-Vernon influence 
functional
of two variables $\Phi(t)\equiv
(\Phi_1^+(t)+\Phi_1^-(t))/2,\chi(t)\equiv \Phi_1^+(t)-\Phi_1^-(t)$,
\begin{eqnarray}
{\cal Z}[\Phi,\chi]=\left\langle\overleftarrow{\rm
T}\exp\left\{\frac{i}{e}\int
dt\bigl[\Phi(t)+{\textstyle\frac{1}{2}}\chi(t)\bigr]\hat{I}(t)\right\}
\hat{\rho}
\overrightarrow{\rm T}\exp\left\{\frac{i}{e}\int
dt\bigl[-\Phi(t)+{\textstyle\frac{1}{2}}\chi(t)\bigr]
\hat{I}(t)\right\}\right\rangle.\label{Zphichi}
\end{eqnarray}
Here, the trace is over the electronic degrees of freedom
specific for
the connector: Thereby they are "traced out"  and never explicitly
enter our field theory. 
The notation $\overrightarrow{T}(\overleftarrow{T})$ denotes time-ordering of the exponentials in ascending (descending) order, these exponents
presenting the quantum evolution of density matrix $\hat{\rho}$
subject to the fields $\Phi,\chi$. 
The easiest way to understand the functional is to expand its log
in terms of $\chi$ at $\chi\to 0$. The coefficents of the expansion present
the cumulants of the time-ordered current operators in the connector
that is subject to the external classical phase $\Phi(t)$. 
In particular, $\ln {\cal Z}(\chi=0)=0$, $\langle I(t) \rangle = \partial
e\ln {\cal Z}/\partial{\chi(t)}$, $\langle\!\langle I(t_1)I(t_2)\rangle\!\rangle =e^2\partial^2 \ln {\cal Z}/\partial{\chi(t_1)}\partial{\chi(t_2)}$ and so on. Therefore, the functional ${\cal Z}$ is nothing but the generating
function of the current fluctuations. For quantum conductors and
slow-varying $\Phi,\chi$ it has been first considered in \cite{Lev93,Naz99}
Now we are ready to build up the quantum description
of the whole circuit. Since the circuit is nothing but
a compound connector, it has to be characterized by
the similar generating function ${\cal Z}_{A+B}$ that depends
on the drop of the fields $\Phi,\chi$ over the circuit.
Such functional is nothing but a path integral convolution of 
the functionals of the separate connectors ${\cal Z}_{A}$
and ${\cal Z}_{B}$,
\begin{equation}
{\cal Z}_{A+B}[\Phi,\chi]=\int{\cal D}\Phi_{1}{\cal D}\chi_{1}\,{\cal
Z}_{A}[\Phi_1,\chi_1]{\cal
Z}_{B}[\Phi-\Phi_{1},\chi-\chi_{1}].\label{concatenation}
\end{equation}
where the path integration measure 
${\cal D}\Phi_1{\cal D}\chi_1 \equiv \prod\limits_t d\Phi(t)d\chi(t)$.  
The overall generating function is the average over fluctuating phases
$\Phi_{1},\chi_{1}$ at the node of the circuit shared by both conductors.
Such convolution law is nothing but the presentation of the
current conservation in the node. One can see it if one substitutes
${\cal Z}_{A+B}$ in the form (\ref{Zphichi}) and carries out the 
integration over $\Phi_1,\chi_1$.

This is the field theory of the simplest single-node circuit. The
extension to a more complicated circuit is straightforward.
The functional is a product of ${\cal Z}$'s of all conductors
integrated over the extra variables $\Phi,\chi$ defined in each
node of the circuit. This functional depends on the phases $\Phi$
applied in each terminal and counting fields $\chi$ defined in terminals
so that it gives statistics of the currents to/from each terminal of the
circuit.

\section{General Properties and Concrete Connectors}
The difference between classical and quantum effects is not
readily manifested in the field theory under consideration. Indeed,
the current fluctuations in the connectors may be of classical as
well as of quantum origin but the presentation of these 
fluctuations is almost the same. For the theory in hand, 
it is constructive
to define the difference between classical and quantum as
the difference between {\it low}-frequency and {\it high}-frequency 
regimes. This is very much like the theory of frequency-dependent
noise: It is known that at sufficiently low frequencies ($\ll k_BT/\hbar$
for equilibrium systems) any
noise can be regarded as classical irrespective of its origin while
quantum mechanics becomes important at higher frequencies.

In general the functional dependence of a 
connector functional ${\cal Z}$ one the phases may be 
complicated and non-local in time. However, one expects on physical grounds
 the non-locality to vanish at sufficiently
low frequencies: The current and its statistics at the time
moment $t$ would only depend on the voltage at the same moment.
Therefore, at 
sufficiently slow realizations of the 
fluctuating phases the
functional ${\cal Z}$ can be expressed in terms of a single function $S(\Phi,\chi)$,
\begin{equation}
\ln {\cal Z}[\Phi(t),\chi(t)]=\int
dt\,S\bigl(\dot{\Phi}(t),\chi(t)\bigr),\label{ZAdef}
\end{equation}
(here we for simplicity specify to non-superconducting systems). 

Let us now concatenate two connectors and try to assess 
the {\it low-frequency} limit of the resulting ${\cal Z}_{A+B}$
taking the path integral in (\ref{Zphichi}).
There can be  two importantly distinct cases. It may be that
the path integral is dominated by low-frequency $\Phi_1,\chi_1$
so that the result is determined by low-frequency limits of
$Z_A,Z_B$. 
Moreover, in this case the path integral can be evaluated
in the saddle-point approximation
 with the result
\begin{equation}
S^{(cl)}_{A+B}(\dot{\Phi},\chi)
=S_{A}(\dot{\Phi}_{s},\chi_{s})+S_{B}(\dot{\Phi}-\dot{\Phi}_{s},
\chi-\chi_{s}).\label{saddlepoint}
\end{equation}
Here $\dot{\Phi}_{s}$ and $\chi_{s}$ stand for the (generally complex) values
of $\dot{\Phi}_{1}$ and $\chi_{1}$ at the saddle point where the derivatives with respect to these phases vanish. Since in field theory
the relevance of saddle-point approximation is normally associated
with classical behavior, we call this classical limit.  
Generally, high-frequency fluctuations of the fields also contribute
to the path integral. The effect of this contribution is that
the actual $S_{A+B}$ deviates from the 
$S^{(cl)}_{A+B}$  evaluated with the saddle-point method. 
Adopting again the common field theory terms, we call the deviation
the {\it quantum correction}. Depending on the connector parameters,
the correction can be vanishingly small or overwhelming.

The next statement might look less obvious: The quantum correction
present the effect of electron-electron interaction in the system 
and in the context of quantum transport is commonly referred to as
(dynamical) Coulomb blockade effect. First puzzled question: Where is
a capacitor of capacitance $C$ providing the 
charging energy $E_C = e^2/2C$ that should accompany any passage about
Coulomb blockade? Well, it always present (Fig. 1b) as a capacitance
between the node and ground. The capacitive connector
contributes to (\ref{Zphichi}) with \cite{Schoen}
\begin{equation}
\ln {\cal Z}_C = \frac{i \hbar}{E_C} \int dt \dot{\Phi}(t)\dot{\chi}(t)    
\end{equation}
This term suppresses the high-frequency fluctuations of the fields
in the node and is usually needed for proper  {\it regularization} of 
the theory since it provides the high-frequency cut-off. 

Common wisdom of quantum transport suggests that Coulomb
interaction is weak provided the typical conductance of the connectors
$G$ exceeds by far the conductance quantum and is dominating otherwise. 
One can see this from the estimation of phase fluctuations
around the saddle point, 
$\langle\!\langle \Phi,\chi \rangle\!\rangle_\omega \simeq 
(i(G/G_Q)\omega + E_C\omega^2/\hbar)^{-1}$. The fluctuation
is $\ll 1$ provided $G \gg G_Q$. We will see that in this regime
the quantum corrections are small although logarithmically diverge
at small frequency. In opposite case our system develops a strong
Coulomb gap and becomes a sort of SET transistor.

We end this Section with concrete examples of connectors. Ohmic
connectors are linear conductors and exhibit Gaussian current 
fluctuations. In terms of Fourier components of the fields,
\begin{equation}
\ln{\cal Z}_{Ohm}= \int \frac{d\omega}{2\pi G_Q} 
\left(\Phi_{\omega} (-i\omega Z^{-1}(\omega))\chi_{-\omega}
- \chi_{\omega}2\omega{\rm Re}(Z^{-1}(\omega))\coth(\hbar\omega/2k_BT)\chi_{-\omega}
\right)
\label{Ohmic}
\end{equation}
$Z(\omega)$ being the frequency-dependent impedance
(resistance) of the connector. Since the action is of
Gaussian type, the field theory is completely
trivial. Any circuits made of Ohmic connectors
are reduced to Ohmic connectors as well.

The Ref. \cite{AES} has provided the connector action for  
 a tunnel connector between the superconducting and/or normal
leads, that appeared to be non-Gaussian even for normal case.
In the simplest low-energy limit such connector is
a Josephson junction characterized by Josephson energy $E_J$
and corresponding action
\begin{equation}
\ln {\cal Z}_J = -i \frac{E_J}{\hbar} \int dt (\cos(2\Phi^+(t)) - \cos(2\Phi^-(t)).  
\end{equation}

An arbitrary coherent quantum connector is characterized
by the set of transmission coefficients $T_n$.
The FCS studies \cite{Lev93} have demonstrated that in low-frequency
limit the functional corresponds to
\begin{equation}
S_{A}(\dot{\Phi},\chi)=\frac{|\dot{\Phi}|}{2\pi}{\cal S}(i\ {\rm sgn}(\dot{\Phi}))\chi),\;\;{\cal
S}(\xi)\equiv \sum_{n=1}^{N}\ln\bigl[1+(e^{\xi}-1)T_{n}\bigr].
\label{Sxi0}
\end{equation}

\section{Saddle point and FCS at current bias}
Let us investigate the saddle point of our field theory
for a quantum conductor (A) of conductance $G$ 
in series with an Ohmic conductor (B)
of low-frequency resistance $Z$. It is instructive to assume that
the Ohmic conductor is noiseless, that is, is kept at vanishing
temperature. Its contribution to the action thus reads (see (\ref{Ohmic}) 
\begin{equation}
S_B(\Phi,\chi)= \frac{i\dot{\Phi}\chi}{2\pi Z G_Q}  
\end{equation}
The subject of interest is how the current fluctuations
produced in the quantum conductor are distributed in the whole circuit.
If the Ohmic resistor is small, $ZG \ll 1$, we have FCS of the quantum
conductor at voltage bias. Increasing $Z$ to $ZG \gg 1$, we
achieve the current bias for this quantum conductor and access the FCS
in this limit.

We apply general Eqs.\ (\ref{ZAdef}) and (\ref{saddlepoint}) to our 
specific circuit that is driven by the voltage source  $V_{0}$. 
To avoid significant quantum corrections 
to this field theory (Coulomb blockade effects \cite{ASI}) we assume that
$|Z(\omega)| G_Q\ll 1$ at frequencies $\hbar\omega\simeq {\rm max}(eV,k_BT)$ 
where the quantum corrections start to form. 
The zero-frequency impedance $Z$ can have any value.

Both the voltage drop $V$ at the quantum conductor and the current $I$
through the conductor fluctuate in time for finite $Z$, with averages
$\bar{I}=V_{0}G(1+ZG)^{-1}$, $\overline{V}=V_{0}(1+ZG)^{-1}$. Voltage
bias corresponds to $ZG\ll 1$ and current bias to $ZG\gg 1$, with
$I_{0}=V_{0}/Z$ the imposed current. There are three characteristic time
scales: $\hbar/{\rm max}(e\overline{V},kT)$, $e/\bar{I}$, and the $RC$-time of the
circuit.  The low-frequency regime on which we concentrate is reached for
current and voltage fluctuations that are slow on any of these time scales.

We seek the cumulant generating function of charge
\begin{equation}
{\cal F}(\xi)=\ln\left(\sum_{q=0}^{\infty}e^{q\xi}P(q)\right)=
\sum_{p=1}^{\infty}\langle\!\langle q^{p}\rangle\!\rangle
\frac{\xi^{p}}{p!},\label{Fxidef}
\end{equation}
where $\langle\!\langle q^{p}\rangle\!\rangle$ is the $p$-th cumulant of the
charge transferred during the time interval $\tau$. It is directly related to the
Keldysh action in the saddle point  (\ref{saddlepoint}) by
\begin{equation}
{\cal F}(\xi)=\tau S_{A+B}(eV_{0}/\hbar,-i\xi).\label{calFdef}
\end{equation}
To characterize the fluctuations of the voltage across the quantum contact,
we will also need the cumulant generating function of phase, ${\cal G}(\xi)$.
We use that in the absence of noise in the Ohmic connector
$V=V_{0}-ZI$. Therefore, ${\cal G}$ is related to ${\cal F}(\xi)$ by
a change of variables. The relation is
\begin{equation}
{\cal G}(\xi)=\sum_{p=1}^{\infty}\langle\!\langle
\phi^{p}\rangle\!\rangle\frac{\xi^{p}}{p!}=\phi_{0}\xi+{\cal
F}(-Z G_Q\xi/2),\label{Gxidef}
\end{equation}
$\phi_0$ being the phase change induced by external voltage in time interval
$\tau$, $\phi_0 \equiv eV_0\tau/2\pi\hbar$.
In the limit $Z\rightarrow 0$ of voltage bias the saddle point of the
Keldysh action is at $\dot{\Phi}_{1}=\dot{\Phi}$, $\chi_{1}=\chi$, and from
Eqs.\ (\ref{saddlepoint}), (\ref{Fxidef}), and (\ref{Gxidef}) one recovers the
results of Ref.\ \cite{Lev93}: The cumulant generating function ${\cal
F}_{0}(\xi)=\tau S_{A}(eV_{0}/\hbar,-i\xi)=\phi_{0}{\cal S}(\xi)$ and the
corresponding probability distribution
\begin{equation}
P_{\phi_{0}}(q)=\lim_{x\rightarrow 0}\frac{1}{q!}\frac{d^{q}}{dx^{q}}
\prod_{n=1}^{N}[1+(x-1)T_{n}]^{\phi_{0}}. \label{Pq0}
\end{equation}
We note that the parameter $\phi_{0}$ is in fact
the number of attempted transmissions
per
channel. The first few cumulants are $\langle
q\rangle_{0}=\phi_{0}G/G_Q$, $\langle\!\langle
q^{2}\rangle\!\rangle_{0}=\phi_{0}\sum_{n}T_{n}(1-T_{n})$, $\langle\!\langle
q^{3}\rangle\!\rangle_{0}=\phi_{0}\sum_{n}T_{n}(1-T_{n})(1-2T_{n})$. In the
single-channel case ($N=1$) the distribution (\ref{Pq0})
has the binomial form (\ref{Pqresult}).

After these preparations we are now ready to generalize all of this to finite
$Z$, and in particular to derive the dual distribution of phase
(\ref{Pphiresult}) under current bias. Calculating saddle-point values
of $\Phi_1,\chi_1$  from Eqs.\
(\ref{saddlepoint}) and (\ref{calFdef}) we observe that ($z\equiv ZG_Q$)
\begin{equation}
{\cal F}(\xi)=\frac{\phi_{0}}{z}[\xi-\sigma(\xi)],\;\;
\sigma+z{\cal S}(\sigma)=\xi.\label{Fxi}
\end{equation}
The implicit function $\sigma(\xi)$ (which is determined from the saddle point of the
action) provides the cumulant generating function of charge ${\cal F}$
for arbitrary series resistance $Z$. One readily checks that
${\cal F}(\xi)\rightarrow\phi_{0}{\cal S}(\xi)$ in the limit $z\rightarrow
0$, as it should.

By expanding Eq.\ (\ref{Fxi}) in powers of $\xi$ we obtain a series of relations
between
the cumulants $\langle\!\langle q^{p}\rangle\!\rangle$ of charge at $Z\neq
0$ and the cumulants $\langle\!\langle q^{p}\rangle\!\rangle_{0}$ at $Z=0$.
For example, to linear order we find $\langle q\rangle=(1+ZG)^{-1}\langle
q\rangle_{0}$, which nothing but the trivial
division of voltage: The mean current
$\bar{I}$ is rescaled by a factor $1+ZG$ coming from the series
resistance. Naively, one may assume that
the same rescaling applies to the fluctuations. Indeed, to second order one finds
$\langle\!\langle q^{2}\rangle\!\rangle=(1+ZG)^{-3}\langle\!\langle
q^{2}\rangle\!\rangle_{0}$, in agreement with elementary circuit theory.

However, if we go to higher cumulants we find that other terms appear, which
can not be incorporated by any rescaling. For example, Eq.\ (\ref{Fxi}) gives
for the third cumulant
\begin{equation}
\langle\!\langle q^{3}\rangle\!\rangle=\frac{\langle\!\langle
q^{3}\rangle\!\rangle_{0}} {(1+ZG)^{4}}- \frac{3ZG}
{(1+ZG)^{5}}\frac{\bigl(\langle\!\langle
q^{2}\rangle\!\rangle_{0}\bigr)^{2}}{\langle
q\rangle_{0}}.\label{thirdcumulant}
\end{equation}
While the first term on the the right-hand-side has the expected scaling form, 
the second term does not. This is generic for $p\geq 3$:  $\langle\!\langle
q^{p}\rangle\!\rangle=(1+ZG)^{-p-1}\langle\!\langle q^{p}\rangle\!\rangle$
plus a non-linear (rational) function of lower cumulants \cite{note3}. All
terms are of the same order of magnitude in $ZG$. 

Turning now to the limit $ZG\rightarrow\infty$ of current bias, we see from
Eq.\ (\ref{Fxi}) that ${\cal F}\rightarrow {\cal F}_{\infty}$ with
\begin{equation}
{\cal F}_{\infty}(\xi)=q_{0}\xi-q_{0}{\cal S}^{\rm
inv}(\xi/z)\label{Finfty}
\end{equation}
defined in terms of the functional inverse ${\cal S}^{\rm inv}$ of ${\cal S}$.
The parameter
$q_{0}=\phi_{0}/z=I_{0}\tau/e$ (which
assumed to be an integer $\gg 1$) is the
number of charges transferred by the bias current $I_{0}$ during the 
time interval $\tau$. Transforming from charge to phase variables by means of Eq.\
(\ref{Gxidef}), we find that ${\cal G}\rightarrow {\cal G}_{\infty}$ with
\begin{equation}
{\cal G}_{\infty}(\xi)=-q_{0}{\cal S}^{\rm inv}(-\xi).\label{Ginfty}
\end{equation}
It is interesting to discuss a single-channel conductor (transmission $T_1$)
separately.
In this case the functional inverse gives the function of a similar form.
Eq.\ (\ref{Ginfty}) reduces to ${\cal
G}_{\infty}(\xi)=-q_{0}\ln[1+T_1^{-1}(e^{-\xi}-1)]$, corresponding to the
Pascal distribution (\ref{Pphiresult}). The first three cumulants are
$\langle\phi\rangle=q_{0}/T_1$,
$\langle\!\langle\phi^{2}\rangle\!\rangle=(q_{0}/T_1^{2})(1-T_1)$,
$\langle\!\langle\phi^{3}\rangle\!\rangle=
(q_{0}/T_1^{3})(1-T_1)(2-T_1)$.

While the charge $Q\equiv qe$ for voltage bias $V_{0}\equiv
h\phi_{0}/e\tau$ is known to have the binomial distribution \cite{Lev93}
\begin{equation}
P_{\phi_{0}}(q)={\phi_{0} \choose q}T_1^{q}(1-T_1)^{\phi_{0}-q},
\label{Pqresult}
\end{equation}
we find that the dual distribution of phase $\Phi\equiv 2\pi\phi$ for current
bias $I_{0}\equiv eq_{0}/\tau$ is the Pascal distribution \cite{note2}
\begin{equation}
P_{q_{0}}(\phi)={\phi-1 \choose q_{0}-1}\Gamma^{q_{0}}
(1-\Gamma)^{\phi-q_{0}}.\label{Pphiresult}
\end{equation}
(Both $q$ and $\phi$ are integers for integer $\phi_{0}$ and $q_{0}$.)

In the more general case not depending on the type of the quantum 
conductor we have found that the distributions of charge and
phase are related in a remarkably simple fashion for $q,\phi\rightarrow\infty$:
\begin{equation}
\ln P_{q}(\phi) = \ln P_{\phi}(q) + {\cal O}(1).\label{PCP}
\end{equation}
(The remainder ${\cal O}(1)$ equals $\ln(q/\phi)$ in the shot-noise limit.)
This formula, which valid with logarithmic accuracy, is a manifestation of charge-phase duality,\cite{Ave91} and holds for any conductors.

The binomial distribution (\ref{Pqresult}) for
voltage bias has the interpretation \cite{Lev93} that electrons hit the barrier
with frequency $eV_{0}/2\pi\hbar$ and are transmitted independently with probability
$T_1$. For current bias the transmission rate is fixed at $I_{0}/e$.
Deviations due to the probabilistic nature of the transmission process are
compensated for by an adjustment of the voltage drop over the barrier. If the
transmission rate is too low, the voltage $V(t)$ rises so that electrons hit
the barrier with higher frequency. The number of transmission attempts
("trials") in a time $\tau$ is given by $(e/2\pi\hbar)\int_{0}^{\tau}V(t)dt\equiv
\phi$. The statistics of the accumulated phase $\phi$ is therefore given by the
statistics of the number of trials needed for $I_{0}\tau/e$ successful
transmission events. This stochastic process has the Pascal distribution
(\ref{Pphiresult}).

For the general multi-channel case a simple expression for
$P_{q_{0}}(\phi)$ can be obtained in the ballistic limit (all $T_{n}$'s close
to 1) and in the tunneling limit (all $T_{n}$'s close to 0). In the ballistic
limit one has ${\cal G}_{\infty}(\xi)=q_{0}\xi/N+q_{0}(N-g)(e^{\xi/N}-1)$,
corresponding to a Poisson distribution in the discrete variable
$N\phi-q_{0}=0,1,2,\ldots$. In the tunneling limit ${\cal
G}_{\infty}(\xi)=-q_{0}\ln(1-\xi/g)$, corresponding to a chi-square
distribution $P_{q_0}(\phi)\propto \phi^{q_{0}-1}e^{-g\phi}$ in the
continuous variable $\phi>0$.  In contrast, the charge distribution
$P_{\phi_{0}}(q)$
is Poissonian both in the tunneling limit (in the variable $q$) and in the
ballistic limit (in the variable $N\phi_{0}-q$).

For large $q_{0}$ and $\phi$, when the discreteness of these variables can be
ignored, we may calculate $P_{q_{0}}(\phi)$ from ${\cal G}_{\infty}(\xi)$ in
saddle-point approximation. If we also calculate $P_{\phi_{0}}(q)$ from ${\cal
F}_{0}(\xi)$ in the same approximation (valid for large $\phi_{0}$ and $q$), we
find that the two distributions have a remarkably similar form:
\begin{eqnarray}
P_{\phi_{0}}(q)&=&N_{\phi_{0}}(q)\exp[\tau\Sigma(2\pi\phi_{0}/\tau,q/\tau)],
\label{Pphi0q}\\
P_{q_{0}}(\phi)&=&N_{q_{0}}(\phi)\exp[\tau\Sigma(2\pi\phi/\tau,q_0/\tau)]
\label{Pq0phi}.
\end{eqnarray}
The same exponential function
\begin{equation}
\Sigma(x,y)=S_{A}(x,-i\xi_{s})-y\xi_{s}\label{Sigmadef}
\end{equation}
appears in both distributions (with $\xi_{s}$ the location of the saddle
point). The pre-exponential functions $N_{\phi_{0}}$ and $N_{q_{0}}$ are
different, determined by the Gaussian integration around the saddle point.
Since these two functions vary only algebraically, rather than exponentially,
we conclude that Eq.\ (\ref{PCP}) holds with the remainder ${\cal
O}(1)=\ln(q/\phi)$ obtained by evaluating $\ln[2\pi(\partial^{2}\Sigma/\partial
x^{2})^{1/2}(\partial^{2}\Sigma/\partial y^{2})^{-1/2}]$ at $x=2\pi\phi/\tau$,
$y=q/\tau$.

\section{Renormalization by Ohmic connector}
We consider the same circuit and turn to analysis of quantum
corrections assuming $Z(\omega)G_Q \ll 1$ in the relevant frequency
region.
We demonstrate that the main
effect of the corrections
can be incorporated into the renormalization of 
energy dependence of the transmission eigenvalues of the quantum connector.
We study this dependence in a non-perturbative limit
to obtain an unexpected result:
owing to accumulation of quantum corrections,
all quantum conductors behave at low energies like
either a single or a double tunnel junction, which divides them into
two broad classes.

It has been shown that at low energy scales the relevant 
part of the electron-electron interaction in mesoscopic conductors
comes from their  electromagnetic environment \cite{OldNazarov,ASI}.
The resulting  dynamical Coulomb blockade has been thoroughly investigated for
tunnel junctions \cite{Grabert2}. 
The measure of the interaction strength is the external 
impedance $Z(\omega)$ at the frequency scale $\Omega = {\rm max}(eV,k_B T)$ determined by either the voltage $V$ at the conductor or its temperature $T$.   If $z \equiv G_Q Z(\Omega)\ll 1$   the interaction is weak, otherwise Coulomb effects strongly suppress  electron
transport.    

A tunnel junction is the simplest quantum conductor
with all transmission eigenvalues $T_n \ll 1$.
Interaction effects for 
general connectors with $T_n \simeq 1$ are difficult to quantify
for arbitrary $z$. 
For $z \ll 1$, one can employ perturbation theory  
to  first order in $z$ \cite{OldNazarov2}. 
The contributions \cite{Zai01,Yey01}
associated the resulting  interaction correction to the conductance 
with  shot noise properties
of the conductor,while the interaction correction to noise has been associated
with the third cumulant of charge transfer \cite{Gal02}.
To sort this out, one shall proceed in the framework
of the field theory outlined where all cumulants are incorporated into
functional dependence of the action on the field $\chi$. 
The recent experiment \cite{Cro02} 
addresses the correction to 
the conductance at arbitrary transmission.

A tunnel junction in the presence of an electromagnetic environment exhibits an
anomalous power-law I-V characteristic, $I(V) \simeq V^{2z+1}$.
The same power law behavior is typical for 
tunnel contacts between one-dimensional interacting electron systems, the so-called Luttinger liquids
\cite{Kane92}. It has  also been found  for contacts with arbitrary
transmission between single-channel conductors in the limit of  weak interactions
\cite{Mat93}. In this case, the interactions have been found to renormalize the
transmission.

In our model of a quantum connector, its transmission probabilities $T_n$ are energy independent in the absence of interactions. We first analyze the quantum correction
to  first order in $z$. We identify an elastic and an inelastic 
contribution. The elastic contribution comes with
a logarithmic factor that diverges at low energies suggesting
that even  weak interactions can suppress  electron transport
at sufficiently low energies. To quantify this we sum up
quantum corrections to the action in all orders in $z$ by a
renormalization group analysis. We show that the result 
is best understood  as
 a renormalization of the transmission eigenvalues similar
to that proposed in \cite{Mat93}. The renormalization brings about
an energy dependence of the transmission eigenvalues according 
to the flow equation 
\begin{equation}
\label{main}
\frac{ d T_n(E)}{d {\rm \ln} E} = 2 z\,  T_n(E) [ 1 - T_n(E)]. 
\end{equation}
To calculate transport properties in the presence of interactions, 
one evaluates $T_n(E)$ at the energy $E \simeq \Omega$.

With  relation  (\ref{main})  
we explore the effect of quantum corrections on 
the distributions of
transmission probabilities for various types of mesoscopic conductors. 
In general, their conductance $G$  and their  noise properties display a complicated
behavior at $z |\ln E| \simeq 1$ that depends on details
of the conductor. However, in the limit of very
low energies $z |\ln E| \gg 1$ we find only two possible scenarios.
The first one is that the conductor behaves like a {\it single} tunnel junction
with $G(V) \simeq V^{2z}$. In the other scenario, the transmission
distribution approaches that of a symmetric  {\it double}
 tunnel junction. 
The conductance scales then  as $G(V) \simeq V^{z}$.
Any  given conductor follows one of the two scenarios. This divides
all mesoscopic conductors into two broad classes.

We still analyze  a simple circuit that
consists of a mesoscopic conductor in series with an external 
resistor $Z(\omega)$ biased with a slow-varying 
voltage source $V_0(t)$ (Fig. 1) but now concentrate
on quantum corrections.

As we have already done, we present the generating function 
${\cal Z}([\chi,\Phi]) $ of the low-frequency current
fluctuations in the circuit is as
a path integral  
over the fields $\Phi_1(t),\chi_1(t)$(Eq. \ref{concatenation}).
It is convenient for us to change the order of 
the connectors so that 
\begin{eqnarray} 
\label{eq:path} 
{\cal Z}(\Phi,\chi) =  \int{
{\cal D} \Phi_1 {\cal D} \chi_1 \exp\left\{ 
\ln {\cal Z}_c \left[\Phi-\Phi_1,\chi-\chi_1\right]
\ln{\cal Z}_{Ohm}
\left[\Phi_1,\chi_1\right]\right\} }
\end{eqnarray} 
where $d\Phi(t)/dt \equiv eV(t)$ and ${\cal Z}_{Ohm}$ is given by
(\ref{Ohmic}).
Let us assume that $\dot{\Phi},\chi$ are slow fields.
In order not to repeat the considerations of the previous
Section, we will simply set  $Z(0)$ to 0. In this case,
the saddle point is trivial: $\Phi_1,\chi_1=0$. In physical
terms, all the voltage drops on the quantum contact.

We start the renormalization  by concentrating on
the "fast" part of the fields $\Phi_1,\chi_1$ and expanding the
action till quadratic terms in these fields. Doing so,
we neglect the time-dependence of slow fields in comparison with that
of fast fields, so the corresponding
part of the action reads 
\begin{equation}
\label{Gaussian-renormalization}
\int dt \int \frac{d\omega}{2\pi}
\phi^\alpha_\omega\left( \frac{\delta^2 S_c(\Phi,\chi)}{
\delta \phi^\alpha_\omega\delta \phi^\beta_{-\omega}}+
M_{Ohm}^{\alpha\beta}(\omega) \right) \phi^\beta_{-\omega}
\end{equation}
where $\alpha,\beta =\pm, \phi^{\pm}_\omega \equiv \Phi_1(\omega)\pm\chi_1(\omega)/2$,
and $M_{Ohm}$ presents (\ref{Ohmic}).
We require that $Z(\omega)G \ll 1$ at any frequency. Under these
conditions, the fluctuations of the fast fields are determined
by the Ohmic term while the part of the action 
that comes from the fluctuation and {\it does} depend on the
slow fields $\dot{\Phi},\chi$ is determined by the
quantum conductor. Indeed, taking the Gaussian integral (\ref{Gaussian-renormalization})
we obtain the contribution to the action 
\begin{equation}
\delta S_c(\Phi(t),\chi(t)) = \int \frac{d\omega}{2\pi}  
\frac{\delta^2 S_c(\Phi(t),\chi(t))}{
\delta \phi^\alpha_\omega)\delta \phi^\beta_{-\omega}} 
\langle\phi^\alpha_\omega \phi^\beta_{-\omega}\rangle =
\int \frac{d\omega}{2\pi}  
\frac{\delta^2 S_c(\Phi(t),\chi(t))}{
\delta \phi^\alpha_\omega\delta \phi^\beta_{-\omega}} 
(M^{-1})^{\alpha\beta}(\omega)
\label{renorm}
\end{equation}
This is the renormalization sought. The correction
to the conductance of the quantum conductor it gives
is of the order of $G z$.

To proceed, we need the action
 of the quantum conductor at fast fields, not just
 at slow ones as given by Eq. \ref{Sxi0}. 
It is expressed in terms of  Keldysh Green functions 
${\check G}_{R,L}$ (the "check" denotes $2 \times 2$
matrices in Keldysh space) of electrons in the  two 
reservoirs adjacent to the conductor \cite{Naz99}.
It takes the form of a trace over frequency and Keldysh indices,
\begin{eqnarray} \label{eq:action} {S}_{\rm c}  = \frac{i}{2}
\sum_n {\rm Tr} \;\ln \left[1 + \frac{T_n}{4} \left( \left\{ {\check G}_{\rm
L} ,{\check G}_{\rm R} \right\} -2 \right) \right]
\end{eqnarray} 
and depends on the set of transmission eigenvalues $T_n$ that characterizes the
conductor. The fields $\phi^{\pm}(t)$ enter the expression as a gauge transform
of $\check G$ in one of the reservoirs,  
\begin{eqnarray}
&&{\check G}_{\rm
R}=   {\check G}^{\rm res} \;\;\; {\rm and} \;\;\;\; {\check G}_{\rm
L}(t,t')= \nonumber \\
&& {\textstyle \left[ \begin{array}{cc}e^{ i\phi^{+}_c(t)} & 0 \cr
0 & e^{i\phi_c^{-}(t)} \end{array}\right] }
{\check G}^{\rm res}(t-t')
{\textstyle \left[ \begin{array}{cc} e^{-i\phi^{+}_c(t')} & 0 \cr
0 & e^{-i\phi^{-}_c(t')} \end{array}\right]} , \nonumber \\
\end{eqnarray}
$\phi_c^{\pm} = \Phi\pm\chi/2-\phi^{pm}$ being the drop of the
phase over the quantum conductor, $G^{\rm res}$ being the equilibrium
 Keldysh Green function  
\begin{equation}
 \label{eq:xiG} 
{\check G}^{\rm res}(\epsilon)= 
\left(
\begin{array}{cc} 1 - 2 f(\epsilon) & 2 f(\epsilon) \\ 2[1 - f(\epsilon)] & 2
f(\epsilon)-1 \end{array} \right),
 \end{equation} 
at a given equilibrium electron distribution function $f(\epsilon)$.

 To zeroth order in $z$ the
fields $\phi^{\pm}(t)$ 
do not fluctuate and are fixed to $eVt \pm \chi/2$.
Substituting this into Eq. (\ref{eq:action}) we recover the slow-field
action (\ref{Sxi0})
\begin{eqnarray} 
\label{eq:zeroorder} 
S^{(0)}(V,\chi)=
\int{\frac{d\epsilon}{2 \hbar \pi} \sum_n \ln\left\{1
+ T_n\left[ (e^{i \chi}-1) f_{\rm
L}(1 - f_{\rm R}) + (e^{-i\chi}-1) f_{\rm R}(1 - f_{\rm L})
\right]
\right\}} 
 \end{eqnarray} 
($f_R \equiv f$ and $f_L(\epsilon)\equiv f(\epsilon - eV)$).  
To assess the renormalization correction,
we expand the non-linear ${S}_{\rm c}$ to second order in the 
fluctuating fields $\phi^{\pm}_\omega$ and use (\ref{renorm}). The expression
for the correction can be presented as 
\begin{eqnarray}
\label{eq:firstorder} 
S^{(1)}(V,\chi)= \int_0^{\infty}{d\omega
\,\frac{{\rm Re}\, z(\omega)}{\omega} \left\{ [2N(\omega)+1] {S}^{(1)}_{\rm el} 
\label{eq:1st1}  
+ N(\omega)
 {S}^{(1)}_{\rm in}(\omega) + [N(\omega)+1]
{S}^{(1)}_{\rm in}(-\omega)\right\}}. 
\end{eqnarray} 
The three terms in square brackets correspond to   {\it elastic} electron transfer,
inelastic  transfer with absorption of energy $\hbar \omega$ from
the environment, and  inelastic electron transfer with emission 
of this energy  respectively. It is crucial  to note that inelastic processes can only occur at frequencies $\omega \leq \Omega$ and that their contribution
to the integral is thus restricted to this frequency range. In contrast, elastic contributions come primarily from  frequencies exceeding
the scale $\Omega$. If $z={\rm const}(\omega)$ for $\omega \leq \Lambda$,
the elastic correction diverges logarithmically, its magnitude being 
$\simeq z \ln \Lambda /\Omega$. This suggests that
i. the elastic correction is more important than the inelastic one
and ii. a small value of $z$ can be compensated for by a large logarithm,
indicating the breakdown of perturbation theory.   
The upper cut-off energy $\Lambda$ is set either by the inverse $RC$-time of  
the environment circuit or 
the Thouless energy of the electrons in the mesoscopic conductor. 

The concrete expression for ${\cal S}^{(1)}_{\rm in}$ reads 
\begin{eqnarray}
\label{eq:1st2}  { S}^{(1)}_{\rm in}(\omega,\chi) = i  \sum_n
\int \frac{d\varepsilon}{2\pi} \, D_n D_n^+\left\{\left.
T_n (f_L-f_L^+)
+ 2 T_n (e^{i\chi}-1)f_L(1-f^+_R)  \right.\right. \nonumber \\
\left.  + 2 T_n^2  (\cos \chi -1)   f_L (1-f_L^+)(f_R^+-f_R)   
 +  T_n D_n + (1-D_n)(1-D_n^+) \right\}  \nonumber \\
 + \left\{  {\rm R} \leftrightarrow {\rm L} ,  {\chi \leftrightarrow -\chi} \right\},
 \end{eqnarray}
where we have introduced the functions
 \begin{eqnarray} \label{eq:1st4} 
D_n= \left\{ 1+ T_n \left[ f_L(1-f_R)(e^{i \chi}-1) 
+f_R(1-f_L)(e^{-i\chi}-1)\right]\right\}^{-1}
 \end{eqnarray} 
and the notation 
\begin{equation} 
f^{+}(\varepsilon)=f(\varepsilon+\omega), \;\;
D_n^+(\varepsilon)=D_n(\varepsilon+\omega). 
\end{equation}
 We do not
analyze ${S}^{(1)}_{\rm in}$ further  and instead turn to the
analysis of the elastic correction. It is important that the explicit 
form of this correction can be presented as 
\begin{eqnarray} \label{eq:zerotemp} 
{ S}^{(1)}_{\rm el}=  \sum_n \delta T_n 
\frac{\partial {S}^{(0)} }{\partial T_n} \ 
{\rm with} \; \delta T_n =  - 2 T_n (1-T_n). 
\end{eqnarray}
This suggests that the main effect of renormalization is {\it to change
the transmission coefficients} $T_n$. It also suggests that 
we can go beyond perturbation theory  by 
a renormalization group analysis that involves the  $T_n$ only.
In such an analysis one concentrates at each renormalization step  on the
"fast" components of $\phi^{\pm}$ with frequencies in a narrow
interval $\delta \omega$ around the running cut-off frequency $E$.
Integrating out these fields one obtains a new action for the slow fields. 
Subsequently one reduces $E$ by $\delta\omega$
and repeats the procedure until the running cut-off approaches $\Omega$.
We find that at each step of renormalization the action indeed retains
the form given by Eq. (\ref{eq:action}) and only the  $T_n$ change,
provided $z \ll {\rm min}\{1,G_Q/G\}$. The resulting energy dependence of the $T_n$ obeys Eq. (\ref{main}).
The approximations that we make in this renormalization procedure amount to a
summation of the leading logarithms in every order of  the  perturbation series.
   
In the rest of the Section 
we analyze the consequences of Eq. (\ref{main}) for various mesoscopic
conductors.
Equation (\ref{main}) can be explicitly integrated  to obtain 
\begin{equation}
\label{eq:tofE}
 T_n(E)  = \frac{\xi T^{\Lambda}_n }
{1 -T^{\Lambda}_n \left(1-\xi\right)}, \;\; 
\xi \equiv \left( \frac{E}{\Lambda}\right)^{2z}
\end{equation}
in terms of the "high energy" (non-interacting) transmission eigenvalues $T^{\Lambda}$.
A mesoscopic conductor containing many transport channels  is most conveniently characterized by
the distribution $\rho_{\Lambda}(T)$ of its transmission eigenvalues \cite{general}.
It follows from Eq. (\ref{eq:tofE}) that the effective transmission distribution at the energy scale $E$ reads
\begin{equation}
\label{eq:rho}
\rho_{E}(T)= \frac{\xi}{[\xi+T(1-\xi)]^2} \rho_{\Lambda}\left( 
\frac{T}{\xi+T(1-\xi)}\right).
\end{equation}
We now analyze its low energy limit $\xi \rightarrow 0$. Any given
transmission eigenvalue will approach zero in this limit. Seemingly
this implies that for any conductor the transmission distribution
 approaches  that of a tunnel junction, 
so that all $T_n \ll 1$.
The overall conductance would be proportional to $\xi$ in accordance
with Ref. \cite{OldNazarov2}. 

Indeed, this is one of the possible scenarios. A remarkable exception
is the case that the non-interacting $\rho_{\Lambda}$ has an inverse square-root
singularity at $T \rightarrow 1$. Many mesoscopic conductors
display this feature, most importantly diffusive ones \cite{general}.
In this case, the low-energy  transmission
distribution approaches a limiting function
\begin{equation}
\label{eq:limiting}
\rho_*(T) \propto \sqrt{\frac{\xi} {T^3(1-T) }}.
\end{equation}
The conductance scales like $\xi^{1/2}$. $\rho_*$
is known to be the transmission distribution of a {\it double} tunnel
junction: two identical tunnel junctions in series \cite{Melsen}.
Indeed, one checks that for a double tunnel  junction the form
of the transmission distribution is unaffected by interactions.
This sets an alternative low-energy scenario. We are not aware of
transmission distributions that would give rise to other scenarios.

We believe that this is an important general result in the theory of
quantum transport and suggest now a qualitative explanation.
The statement is  that the conductance of a phase-coherent
conductor at low voltage and temperature $\Omega \ll \Lambda$
asymptotically obeys  a power law with an exponent that  generically takes two values, 
\begin{equation}
\label{eq:scaling}
G \propto \left(\frac{\Omega}{\Lambda}\right)^{2z} ,\;\; {\rm or} \;\;\;G \propto \left(\frac{\Omega}{\Lambda}\right)^{z}.
\end{equation}   
For tunneling electrons the exponent is $2z$. 
An electron traverses the conductor in a single leap.
The second  possible exponent  
$z$  has been  discussed in the  literature as well, 
in connection with  { \it resonant} tunneling through a {\it double} tunnel barrier in the presence of
interactions \cite{Kane92}. This resonant tunneling takes place via intermediate
discrete states contained between the two tunnel barriers. 
  The halved exponent $\alpha=z$ occurs in  the regime of the so-called {\it successive} 
electron tunneling. In this case, the  electron first jumps over
one of the barriers ending up in a discrete state. Only in a second jump over
the  second barrier the charge transfer is completed.
Since it takes two jumps to transfer a charge, 
the electron feels only half the counter voltage due to interactions with electrons in the environmental impedance $Z$ at each hop. Consequently, the exponent  at each jump takes  half the  value for direct tunneling.
Our results strongly suggest that this transport mechanism
is not restricted to resonant tunneling systems, or, in other words, that   resonant tunneling can occur in  systems of 
a more generic nature than generally believed. 
As far as transport is concerned, a mesoscopic conductor is characterized by its scattering
matrix regardless of the details of its inner structure.
 In this approach  it is not even obvious that 
the conductor can accommodate  discrete   states.
Nevertheless,  the transmission distribution of this scattering matrix does depend on the internal structure of the conductor.  The inverse square root singularity of this distribution at $T\to1$ for a double tunnel barrier is   due to   the formation of Fabry-Perot  resonances  between the two  barriers. Probably similar resonances are at the origin of the same singularity for  more complicated mesoscopic conductors with multiple scattering. They are then  the intermediate
discrete states that give rise to the modified scaling of the conductance in  presence of interactions. One may speculate that in diffusive conductors these resonances are the so-called ''nearly localized states'' found in  \cite{Khm95}. 

From equation (\ref{eq:tofE}) one concludes
that the resonant tunneling scaling holds only if $G(E) \gg G_Q$
so that many transport channels  contribute to the conductance.
At sufficiently small energies, $G(E)$ becomes of the order of $G_Q$.
All  transmission eigenvalues are then small and the conductance crosses over to  the  tunneling scaling.

\section{Renormalization by quantum connectors}
\begin{figure}[t]
\begin{center}
\includegraphics[width=3in]{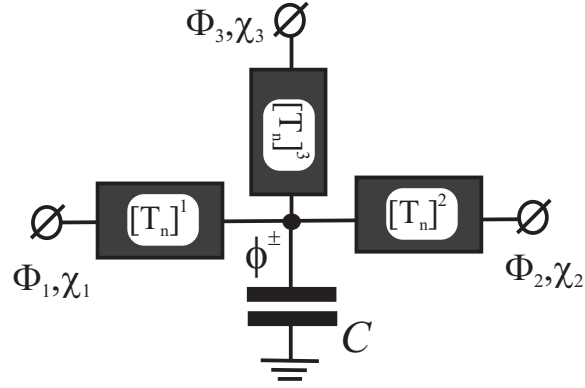}
\caption{Coulomb island setup: A single node connected
to $M=3$ terminals by quantum connectors. The field theory
is for the fields $\phi^{\pm}(t)$.}
\label{Coulomb-island}
\end{center}
\end{figure}

One could wonder about the generality of the results
obtained in the previous Section and expressed by Eq. \ref{main}
Indeed, it has been proven under rather restrictive assumptions
of an Ohmic connector of negligible resistance. Here, we consider
a more general model of several quantum connectors coming together
in a single node (Fig. \ref{Coulomb-island}). Each
connector labeled by $k$ is characterized by
the set of the transmission eigenvalues $T_n^{[k]}$.
In traditional Coulomb blockade situation ($G \ll G_Q$)
this setup is called Coulomb island or SET transistor \cite{ASI}
and is seen very different from the junction-in-the environment 
setup considered in the previous Section.
However, we show that in the limit of large conductance $G \gg G_Q$
the Coulomb island is governed by very 
similar renormalization equations:
\begin{equation}
 \frac{d\, T_n^{[k]}}{d\ln E} = 
\frac{2\, T_n^{[k]}(1- T_n^{[k]})}{\sum_{n,k}  T_n^{[k]}}.
\label{RGEquation}
\end{equation} 
It looks like each quantum channel sees all others
as an "environment" characterized by the effective
island conductance $g=\sum_{n,k}  T_n^{[k]}$. If we
use $z=g^{-1}$, Eqs. \ref{RGEquation} and \ref{main} are
identical. The difference is that $g$ by itself is subject
to renormalization. 

This results in very different 
low-energy behavior.
In contrast to the considerations of the  previous Section,
 the renormalization of all transmission eigenvalues
may break down at finite energy --- effective Coulomb gap --- 
$\widetilde E_C \propto   g_0 E_C e^{-\alpha g_0}$, 
$\alpha$ being a numerical factor depending on the details
of the initial transmission distribution, $g_0$ is the island conductance
at high energies $> E_C g_0$.
Remarkably, $\widetilde E_C$ coincides 
with the effective charging energy evaluated with
instanton technique.\cite{Nazarov} However, the renormalization
stops at the effective Thouless energy  
$E_{\rm Th} \sim G(E) \delta/G_Q$, 
$\delta$ being mean level spacing in the island. 
This gives rise to 
{\it two} distinct 
scenarios at low energy.
If $g_0 > \alpha^{-1}\ln(E_C/\delta)$, Coulomb blockade does not
occur with zero-bias conductance being saturated 
at the value $G(E_{\rm Th}) \gg G_Q$.
Alternatively, 
$G(0)\approx 0$ and $\widetilde E_C$ defines the Coulomb gap.  
 
Let us give the details of the model in use. 
The Coulomb island is characterized by two parameters:
charging energy $E_C$ and 
the mean level spacing $\delta$, $E_C \gg \delta$. \cite{Remark}
The island is connected to $M \ge 2$ external leads
by means of $M$ arbitrary quantum connectors (Fig. \ref{Coulomb-island})  
 characterized by the set
of transmission eigenvalues $T_n^{[i]}$.  
We assume that the island is strongly coupled to the leads,
$g_0 = \sum_{n,m} T_n^{[i]} \gg 1$.
Our goal is to evaluate the 
functional ${\cal Z}$ for the whole circuit
that now depends on voltages and counting fields in each
terminal, ${\cal Z}([V_i,\chi_i]))$.
To evaluate this for the Coulomb island, we have extended
the semiclassical approach for the FCS of the non-interacting
electrons~\cite{NazBag}. 
The node houses a dynamical phase variable $\phi(t)$~\cite{Schoen} 
, its time derivative, $\dot\phi(t)/e$,
presents the fluctuating electrostatic potential of the island.
According to the rules of our field theory,
the functional is represented in the form of a 
real-time path integral over the fields $\phi^{\pm}(t)$ residing
at two branches of the Keldysh contour
\begin{eqnarray}
&&{\cal Z}(\{V_i,\chi_i\}) = \int D\phi_{\pm}(t)\exp\Big\{
\frac{i}{2}E_C^{-1}\int\limits_{-\infty}^{+\infty} d\,t
((\dot\phi^+)^2-(\dot\phi^-)^2) \nonumber \\
&& - \sum_k S^{[k]}_{\rm con}\bigr(\{\hat G,\hat G_k^\chi\}\bigl) \,-\,
 i\pi\delta^{-1}{\rm Tr}\{(i\partial_t - \dot\Phi)\hat G\} 
\Big\} \label{Model}  
\end{eqnarray}
Here  
$
\hat\Phi = \left(
\begin{array}{cc}
  \phi^+(t) & 0 \\
   0  &  \phi_-(t)
\end{array}
\right)
$ is the matrix in Keldysh space, $2\times 2$ matrix $\hat G(t_1,t_2)$
presents  the electron Green function in the island
that implicitly depends
on $\phi^{\pm}(t)$. 
The trace operation includes the summation
over Keldysh indices and the integration in time. 
The contribution of each connector $S^{[k]}_{\rm con}$ 
has a form (\ref{eq:action}) 
\begin{equation}
 S_{\rm con}^{[k]} = -\frac{1}{2}\sum_{n} {\rm Tr} \ln\left[1+ \frac{1}{4}T_n^{[k]}
(\{\hat G,\hat G_k^\chi\} - 2)\right] 
\label{Sk}
\end{equation} 
$\{\hat G,\hat G_k^\chi\}$ denoting the
anticommutator of the Green functions with 
respect to both Keldysh and time
indices. The Green functions in the leads $\hat G_k(\chi)$  
are obtained by $\chi$-dependent gauge
transformation ~\cite{Wolfgang} 
of the equilibrium Green functions in the reservoir $k$, 
$\hat G_k^{[0]}$,
$
\hat G_k^\chi(\epsilon) = \exp (i \chi_k \bar \tau_3/2) \hat G^{(0)}_k(\epsilon) 
\exp (-i \chi_k \bar \tau_3/2)
\label{boundary}
$,
where $\hat G_k^{[0]}$ are given by
$
\bar G_k^{[0]} = \left(
\begin{array}{cc}
 1-2f_k & -2 f_k \\
 -2(1-f_k) & 2f_k-1
\end{array} \right) 
$.
Here  $f_k(\varepsilon)$ 
presents the electron distribution function in the $k$-th reservoir.
The expression (\ref{Sk})
is valid under assumption of instantaneous electron transfer
via a connector, thus corresponding to energy-independent $T^{[k]}_n$. 

In order to find $\hat G(t_1,t_2)$ at given $\phi_{\pm}(t)$,
we minimize the action with respect to all $\hat G(t_1,t_2)$ 
subject to the constrain 
$\hat G \circ \hat G = \delta(t_1-t_2)$.
This yields the saddle point equation for  
$\hat G(t_1,t_2)$:
\begin{equation}
 \sum_{n,\,k}
  \frac{T^{[\,k]}_n [\hat G_k^\chi,\hat G ]}{
    4+T^{[\,k]}_n\left(\{\hat G_k^\chi, \hat G\} - 2 \right)} = 
  i\pi\delta^{-1}[\,i\partial_t - \dot\Phi, \hat G ] \label{Saddle}
\end{equation}
where $[..\,,..]$ denotes the commutator in the Keldysh-time space. 
This relation
expresses $\hat G(t_1,t_2)\equiv \hat G(t_1,t_2; 
[\phi^{\pm}(t)])$  
via the reservoir Green functions $\hat G^{[k]}$. 
This circuit theory relation is similar to obtained in \cite{NazBag}.
It disregards the mesoscopic fluctuations, since those
lead to
corrections of the order of $\sim 1/g_0$ at all energies, 
whereas the interaction corrections are of the order of $\sim 1/g_0 \ln(E)$
tending to diverge at small energies. 
If $\phi_{\pm}(t)=0$, 
Eq.~(\ref{Saddle}) separates in energy representation
and coincides with that of Ref. \cite{NazBag}.

This sets the model.
We start the analysis of the model with
perturbation theory in  $\phi_{\pm}$
around the semiclassical saddle point $\hat G(t_1,t_2)=\hat G_0$,
$\phi_{\pm}(t)=0$. 
The phase fluctuations are
small, $\delta \phi^2 \sim 1/g_0$, so we 
keep only  quadratic 
terms to the action ~(\ref{Model}).
The resulting Gaussian path integral over $\phi_{\pm}$ 
can be readily done.
This procedure is equivalent to the summation
of all one-loop diagrams of the conventional perturbation theory, i.e.  
to the  "random-phase approximation" (RPA).

We restrict ourselves
 to the most interesting 
low voltage/temperature limit, $\max\{eV,kT\}\ll g_0 E_C$.
In this limit, we evaluate the interaction correction to the CGF
with the logarithmic accuracy.
It reads
\begin{eqnarray}
&&\Delta S_\chi = \frac{t_0}{g_0}\ln\left(\frac{g_0 E_C}{\max\{eV,kT\}}\right) \times 
\label{S2} \\
&&\int \frac{d\varepsilon}{2\pi} \sum_{n,\,k}
\frac{2T^{[\,k]}_n(1-T^{[\,k]}_n)\bigl(\{\hat G_k^\chi,\hat G_0 \}-2\bigr)}{
    4+T^{[\,k]}_n\bigl(\{\hat G_k^\chi, \hat G_0\} - 2 \bigr)} \nonumber
\end{eqnarray}
provided $\max\{eV,kT\} > E_{\rm Th}$, 
where $E_{\rm Th}=g_0\delta$ is the Thouless energy of the island. 
In the opposite case,
$\max\{eV,kT\}< E_{\rm Th}$, 
the voltage/temperature should be replaced
with $E_{\rm Th}$.
Note, that the correction (\ref{S2}) is contributed by only 
virtual inelastic processes that change the probabilities
of real elastic scatterings.

For simplicity, we consider the shot-noise limit
$eV\gg kT$ only.  
Then the magnitude of the correction  
shall be compared with the zero-order CGF $S^{[\,0]}_\chi\sim t_0 eV g_0$. 
This implies that the perturbative RPA result~(\ref{S2}) 
is applicable only if 
$\displaystyle g_0^{-1}\ln\left({g_0 E_C}/{eV}\right)\ll 1$.
At lower voltages $\Delta S_\chi$ 
logarithmically diverges. This indicates that we should
proceed with a renormalization group (RG) analysis.
   
We perform the RG analysis of the action~(\ref{Model}) 
along the lines of the previous section
 decomposing $\phi^{\pm}(t)$ onto the fast 
and slow parts . 
On each step of RG procedure we
eliminate the fast degrees of freedom in
the energy range $E-\delta E<\omega<E$ 
to obtain new action $S_{E -\delta E} [\phi_s]$,
$E$ being the current ultraviolet cutoff.
Our key result is that 
the change in the action at each step of RG procedure can be
presented as a change 
of transmission eigenvalues
$T_n^{[k]}$.
Therefore, the RG equations can be written directly for
transmission eigenvalues and take a simple form (\ref{RGEquation}).
The equations are to be solved 
with initial conditions at the upper cutoff energy 
$E = g_0 E_C$, those are given by "bare" transmission
eigenvalues $T_n^{[k]}(E = g_0 E_C) =T_n^{[k]}$. 
The RG equations resemble 
those for the transmission coefficient 
for a scatterer in the weakly interacting one-dimensional electron 
gas~\cite{Mat93} and for a single multi-channel scatterer in the
electromagnetic environment \cite{Kindermann}.
The effective impedance $Z$ is just replaced by
inverse conductance of the island to all
reservoirs, $G(E)= G_Q\sum_{n,k} T_n^{[k]}(E)$.
The important difference is that this conductance is itself
subject to renormalization. 
The difference becomes most evident in the case when 
all  contacts are tunnel junctions, $T_n^{[k]}\ll 1$. In this
case, one can sum up over $k,n$ in 
Eqs. (\ref{RGEquation}) to obtain the RG for the
conductance only : $dG/d\ln E = 2 G_Q$. 
This renormalization law \cite{Efetov} was recently
applied to conductance of granular metals. 
The Eqs.~(\ref{RGEquation}) 
could be also derived in the framework of functional 
RG approach to $\sigma$-model of disordered metal.~\cite{Feigelman}.

We solve the RG Eqs.~(\ref{RGEquation}) 
in  general case to obtain 
\begin{eqnarray}
  T_n^{[k]}(E)
&=&T_n^{[k]}y/\left(1-T_n^{[k]}(1-y)\right), 
\label{Solution}\\
\ln({g_0 E_C}/E) &=& -\frac{1}{2}\sum_{n,\,k}
\ln(1-T_n^{[k]}(1-y))
\end{eqnarray}
The first equation gives the renormalized transmission eigenvalues
at a given value $E$ of the upper cutoff in terms
of variable $y(E)$, $0\leq y \leq 1$. 
The second equation implicitly expresses $y(E)$.

\begin{figure}[t]
\centerline{\includegraphics[width=4.5in]{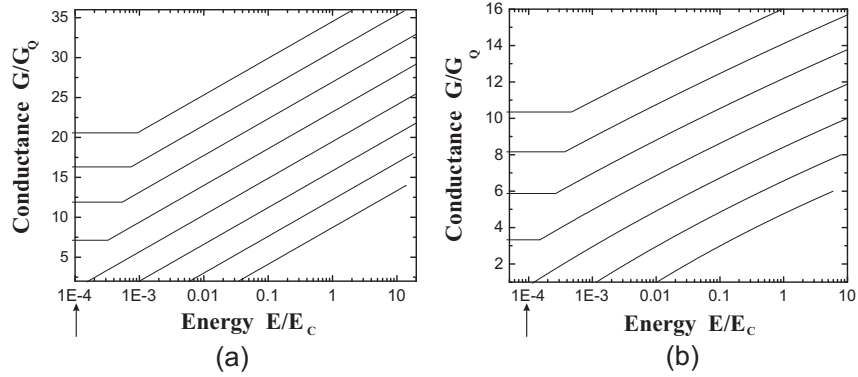}}
\caption{
The total conductance of the Coulomb island 
versus the energy: two scenarios. 
We assume $\ln(E_c/\delta)=10.0$. 
Arrows show the energy scale $\sim \delta$.
Pane (a): tunnel connectors, $g_0$ changes from $42$ (upper curve)
to $14$ (lowermost curve) with the step $4$.
Pane (b): diffusive connectors, $g_0$ changes from $18$ to $6$ with the step $2$.
The conductance either hits $0$ manifesting the Coulomb gap
or saturates at finite value.
} 
\end{figure}

We note that the energy dependence of transmission coefficients
induced by interaction is very weak provided $G(E) \gg G_Q$:
If energy is changed by a factor of two, 
the conductance is changed by $\sim G_Q$. 
To use the equations 
for evaluation of FCS at given voltages $V^{[k]}$ of the leads,
one takes $T_n^{[k]}(E)$ 
at upper cutoff  $E = {\rm max}_k(V^{[k]})$, and further disregards
their energy dependence.
Then one can follow the lines of Ref.~\cite{NazBag}:
It is convenient to introduce the 
function $S^{[k]}(x) = -\sum_n\ln[1+\frac{1}{2}T_n^{[k]}(x-1)]$
to incorporate all required information about transmission eigenvalues.  
The renormalization of $S^{[k]}$ in terms 
of $y$ is especially simple: 
$S^{[k]}(x,y) = S^{[k]}((x +1)y-1)-S^{[k]}(2y-1)$.
From this one readily finds the conductance of each scatterer, 
$G^{[k]}(y) = 2 G_Q \partial S^{[k]}/\partial x(1,y)$, 
as well as  the renormalized
transmission distribution  
$T^2 \rho^{[k]}(T,y) = 
({2}/{\pi}){\rm Im}  \{\partial S^{[k]}/\partial x\,(1-2/T - i0,y)\}$.

The RG equations (1) have a fixed point at $T^{[k]}_{n}=0,y=0$
that occur at finite energy
\begin{equation}
E=\widetilde E_C = g_0 E_C\prod_{k,n}(1-T^{[k]}_n)^{1/2} 
\end{equation}
This indicates the breakdown of RG and formation of Coulomb blockade
with the exponentially small gap $\widetilde E_C$.
The same energy scale was obtained 
from equilibrium instanton calculation of Ref. \cite{Nazarov}.
For a field theory, one generally expects different physics
and different energy scales for instantons and 
perturbative RG. The fact that these scales are the same
shows a hidden symmetry of the model which is yet to understand.

Alternative low-energy behavior is realized if the current cut-off
reaches $E_{\rm Th}= G(E)\delta/G_Q$. (Fig. 3) 
The log renormalization of the transmission
eigenvalues stops at this point and their values saturate.
We thus predict a sharp crossover between the two alternative
scenarios, that occur at value of $g_0=g_c$ corresponding to 
$\widetilde E_C \simeq \delta$. This value equals $g_c = \alpha^{-1}\ln(E_C/\delta)$,
where $\alpha = \frac{1}{2}g_0^{-1}\sum_{n,k}\ln(1-T_n^{[k]})$,  
and depends on
transmission distribution of all connectors. 
If all connectors are tunnel junctions, $\alpha_T = 2$. For diffusive
connectors, $\alpha_D = \pi^2/8$ and the energy dependence of the total conductivity is
given by  $g_D(V)\sim g_0\sqrt{\xi}\,{\rm ctg}{\sqrt{\xi}}$, 
$\xi \equiv 2g_0^{-1}\ln(g_0 E_C/eV)$. (Fig. 3)    



\begin{acknowledgement}
The author gladly acknowledges collaborations 
with C. W. J. Beenakker, M. Kindermann and D. Bagrets
on the questions discussed. 
\end{acknowledgement}



\begin{thebibliography}{10}
\bibitem{note3} We record the result for the fourth cumulant, obtained by
expansion of Eq.\ (\protect\ref{Fxi}) to order $\xi^{4}$: $\langle\!\langle
q^{4}\rangle\!\rangle=(1+ZG)^{-5}\mu_{4}-10z
g(1+ZG)^{-6}\mu_{2}\mu_{3}/\mu_{1}+15(ZG)^{2}(1+ZG)^{-7}
\mu_{2}^{3}/\mu_{1}^{2}$, where we have abbreviated
$\langle\!\langle q^{p}\rangle\!\rangle_{0}=\mu_{p}$.
\bibitem{note2} The Pascal distribution $P(m)={m-1\choose
M-1}\Gamma^{M}(1-\Gamma)^{m-M}$ is also called the ``binomial waiting-time
distribution'', since it gives the probability of the number $m$ of independent
trials (with success probability $\Gamma$) that one has to wait until the
$M$-th success. It is related to the negative-binomial distribution
$P(n)={n+M-1\choose n}\Gamma^{M}(1-\Gamma)^{n}$ by the displacement $n=m-M$.
\bibitem{Remark} Strictly speaking, a purely electric
field theory corresponds to $\delta_S =0$. A finite
$\delta_S$ enables electrons rather than electricity
to escape the node and may provide coherent electron
transfer through two connectors.
\end{thebibliography}
\end{document}